\documentclass[11pt]{article}
\usepackage{epsfig}
\usepackage{graphicx}
\begin{document}
\pagestyle{myheadings}
\thispagestyle{empty}
\normalsize
\title{{\large \bf Evolution of Supermassive Black Holes from Cosmological Simulations}}

\author{Ch. Filloux$^1$, F. Durier$^{1,2}$, J.A. de Freitas Pacheco$^1$~ and~ J. Silk$^3$\\
\small
$^1$Observatoire de la C\^ote d'Azur, 06304 Nice Cedex 4, France\\
\small
$^2$present address: Max Planck Institute for Extraterrestrial Physics\\
\small
$^3$Oxford Astrophysics, Keble Road, Oxford OX1 3RH England}

\date {\today}
\maketitle       
\vspace{0.5cm}

\begin{abstract}

The correlations between the mass of supermassive black holes and properties of their host galaxies 
are investigated through cosmological simulations. Black holes grow from seeds of $100~M_{\odot}$ inserted into density
peaks present in the redshift range 12-15. Seeds grow essentially by accreting matter from a nuclear disk and also
by coalescences resulting from merger episodes. At $z=0$, our simulations reproduce the black hole mass
function and the correlations of the black hole mass both with stellar velocity dispersion and 
host dark halo mass. Moreover, the evolution of the black hole mass density derived from the present
simulations agrees with that derived from the bolometric luminosity of quasars, indicating that the average
accretion history of seeds is adequately reproduced . However, our simulations are unable to form black holes with masses 
above $10^9~M_{\odot}$ at $z\sim 6$, whose existence is inferred from the bright quasars detected  by the Sloan
survey in this redshift range.

\end{abstract}


\section{Introduction}

Observations indicate that the majority of galaxies (and possibly all of them) host a 
supermassive black hole (SMBH) in their centers$^{1-3}.$  The most dramatic example
concerns the Milky Way, where the orbital motion of stars around Sgr A$^*$ can be
explained by the presence of a dark massive object$^4$, with a mass 
around $(3-4) \times 10^6~M_{\odot}$. Black hole
masses are well correlated with such properties of the host galaxy as the projected central
velocity dispersion$^{5,6}$, the stellar mass or the luminosity of the bulge component$^{7,8}$.
Although obtained indirectly, other investigations also suggest a correlation with the 
mass of the dark halo associated to the host galaxy$^{9,10}$, an indication
that  black hole growth follows the same evolutionary path as hierarchical structure formation.

The mechanism (or mechanisms) responsible for establishing the aforementioned relations is
(are) not yet well determined but several scenarios have been put forward in the
literature. There is now a general consensus that SMBHs are the consequence of 
growth mainly by accretion processes, including some contribution from
coalescences resulting from merger episodes of ``seeds" originating from the first stars. 
This picture is consistent with the fact that
the present BH mass density agrees with that derived from the bolometric luminosity function
of quasars under the assumption that the radiated energy comes from an accretion disk around
the black hole$^{11,12}$. In most of the  scenarios considered, the growth of the seeds is
self-regulated by feedback effects due to outflows or UV-radiation associated with the
accretion process itself. These feedback mechanisms also affect  the star formation activity,
leading to a coeval evolution of the black hole and the host galaxy.

Recently, a series of investigations based on analytic or semi-analytic models has
been performed  that aims to study the growth history of black holes and their influence on the
evolution of the host galaxies$^{13-15}$. Despite correctly including the merger trees of halos, these
approaches do not adequately describe all of the hydrodynamic processes related to  gas accretion
and ejection by  feedback due to supernovae and AGNs. Thus, cosmological simulations seem to be an
indispensable tool for investigating the origin and the evolution of SMBHs. Past studies
based on cosmological simulations have found a correlation between the gas mass inside a simulated
galaxy and a suitable defined circular velocity$^{16}$, whose slope at $z=1$ is quite close to the slope
of the $M_{BH}-\sigma$ relation observed today ($z=0$). According to the authors$^{16}$, this relation
is established when the BH growth is limited by the available gas, in competition with star formation
and feedback processes. Moreover, at $z=1$ the SMBH masses are related to the host dark halo masses according to  $M_{BH} \propto M_h^{4/3}$. Other authors$^{17}$ have also compared the black hole growth with that of the host dark matter halo, concluding that despite the existence of a correlation between masses at low redshift, there are deviations from the simple hierarchical picture at early evolutionary phases. In particular, in the redshift range $2 \leq z \leq 6$, BHs grow at a faster rate than their host halos. Cosmological simulations in which seeds of $10^5~M_{\odot}$ grow by spherical accretion were performed in Ref. 18, from which a robust correlation between the SMBH mass with either the stellar velocity dispersion or the stellar mass was found at $z=1$.

In this paper, we report results derived from cosmological simulations in which both the evolution of
galaxies and black holes are followed. In the present approach, depending on the gas angular
momentum, seeds may accrete either in the ``disk" mode (to be detailed later) or in the
``spherical" mode. The former  case, analogous to cold accretion, permits seeds of $100~M_{\odot}$ to grow by several orders of magnitude in a few Gyr, something that is not possible if accretion occurs only in the ``spherical" mode, unless a supercritical Eddington regime is invoked. The present simulations adequately reproduce the relations between the BH mass and properties of the host galaxy, describing as well the resulting evolution of the BH mass density. In Section 2 the details of the simulations are given, the most important results are presented in Section 3,  and finally, in Section 4,  our main conclusions are summarized.

\section{The simulations}

The simulations were performed by using the parallel TreePM-SPH code GADGET-2 in a formulation, despite the use of fully adaptive smoothed particle hydrodynamics (SPH), that conserves energy and 
entropy$^{19}$. Different physical mechanisms affecting the gas properties were introduced, such as
ionization equilibrium of hydrogen and helium, taking into account collisional, radiative processes and
the contribution of the ionizing radiation background; cooling, including free-free transitions, radiative 
recombinations, collisional excitation of molecular hydrogen,
fine-structure levels of trace elements and Compton interactions with CMB photons; local heating
by the UV-radiation of newly formed stars, mechanical energy injected by type II and type Ia supernovae
as well as by AGNs and turbulent diffusion of elements produced by supernova explosions. Astration
processes were also taken into account by computing the return of mass into the interstellar
medium under the assumption that the initial mass function (IMF) of stars with $[Z/H] < -2$ is of the
form $\xi(m) \propto m^{-2}$, while more metal-rich stars have an IMF of the form $\xi(m) \propto m^{-2.35}$. Lower and upper limits for stellar masses were taken respectively equal to $0.1~M_{\odot}$ and $80~M_{\odot}$.

In our simulations, the $\Lambda$CDM cosmology was adopted, characterized by a Hubble parameter $h=0.7$ (in units of $H_0=100~km/s/Mpc$), by a ``vacuum" density parameter $\Omega_{\Lambda}=0.7$ and by a total matter density parameter $\Omega_m=0.3$. The fraction of baryonic matter corresponds to $h^2\Omega_b=0.0224$ and the normalization of the matter density fluctuation spectrum was taken to be $\sigma_8=0.9$. Initial conditions were established according to the algorithm COSMICS$^{20}$ and all simulations were performed in the redshift range $60 \geq z \geq 0$.

\subsection{Black holes}

BHs are represented by collisionless particles that can grow in mass, according to specific rules that mimic accretion or merging with other BHs. A coalescence is supposed to occur whenever the two BHs come within a distance less than the mean inter-particle separation and their total energy is negative.

Seeds are supposed to have originated from the first (very massive) stars and are assumed to have a mass of $100~M_{\odot}$. An auxiliary algorithm finds potential minima where seeds are inserted in the redshift interval 15-20. During the time interval in which the BH mass is much smaller than the mass resolution, we artificially  maintain the seeds at the minima of the halo gravitational potential, since during this phase, dynamical friction effects on the BH are not correctly described by the code.

The simulation of the accretion process requires substantial simplification since it occurs on unresolved
physical scales. If the BH is at rest with respect to the gas, the inflow geometry is probably spherical
and almost adiabatic, with the accretion rate given by the well known equation
\begin{equation}
\frac{dM}{dt}=\frac{\pi}{4}\lambda(\gamma)r^2_g\left(\frac{c^4}{a^3_{\infty}}\right)\rho_{\infty}
\end{equation}
where $\lambda(\gamma)=\left[2/(5-3\gamma)\right]^{(5-3\gamma)/2(\gamma-1)}$, $\gamma$ is the
adiabatic index, $r_g$ is the gravitational
(or Schwarzschild) radius, $\rho_{\infty}$ and $a_{\infty}$ are respectively the density and the sound
velocity of the gas far away from the radius of influence of the BH. If the BH is not at rest but moving with a velocity $V$ with respect
to the gas, then the sound velocity in the denominator of the equation above should be replaced by
$a_{\infty}^2 \rightarrow V^2+a_{\infty}^2$. On the other hand, after a merger event, numerical simulations indicate that most of the gas will have settled into a central self-gravitating disk$^{21}$, which will be able to feed the central BH. Since the present simulations are unable to describe the physics of these nuclear disks with dimensions of about 500 pc, the accretion rate must be estimated under simplifying assumptions. We assume that during a single time step, the properties of the disk do not change appreciably or, in other words, the disk is steady but its properties are continuously updated. Under these conditions, the accretion rate is given by
\begin{equation}
\frac{dM}{dt}=\frac{6\pi}{{\cal R}}\frac{c_sV^2_{\phi}}{QG}
\end{equation}
where the free parameters of our model are ${\cal R}$, the critical Reynolds number of the flow and $Q$,
the Toomre parameter, which guarantees the disk stability. $V_{\phi}$ and $c_s$ are respectively the 
tangential and the sound velocities of the gas and $G$ is the gravitational constant. This equation is 
applied whenever the angular momentum of
the ``gas particles" representing the disk is lower than the corresponding angular momentum of circular
orbits at the considered particle position.

\subsection{Parameters of the simulations}

A series of runs aiming to test the different parameters and resolution effects were performed at the Centre of Numerical Computation of the Observatoire de la C\^ote d'Azur (SIGAMM) whose results will be reported elsewhere. Here, our analysis is focused only on three simulations performed with the same mechanical energy input  by supernovae ($5\times 10^{49}$ erg for type Ia and $3\times 10^{49}$ erg for type II) but with two different mass resolutions, i.e., $2\times 160^3$ and $2\times 192^3$ particles, corresponding for a cubic volume of size of $50h^{-1}~Mpc$ and  an initial mass resolution respectively of $5.35\times 10^8~M_{\odot}$ and $3.09\times 10^8~M_{\odot}$ for gas/stellar particles. In runs 192disk and 160disk, the injected energy rate by AGNs is about 5\% of the gravitational energy rate released by the accretion process whereas in run 160kerr, the injected energy rate is based on the Blandford \& Znajek mechanism$^{22}$. In all these simulations, the Toomre parameter was taken as $Q=100$, whereas the critical Reynolds number was taken equal to 20,000 in runs with $2\times 160^3$ and equal to 8,000 in the run with $2\times 192^3$ particles.

\section{Results} 

Percolation algorithms ({\it FOF and SubFind}) identify the main structures and then, by an iterative procedure, unbound particles are eliminated from the selected objects. Thus, all the objects in the catalogs resulting from each run are gravitationally bound. The resulting dynamical and photometric properties of galaxies are calculated under the assumption that each stellar particle can be associated with a stellar cluster or a ``single population object", with a well defined mass, age and chemical abundance. Since our simulations do not have enough resolution to distinguish the morphology of galaxies, a colour criterion was adopted: {\it red galaxies}, including all the objects with integrated
colours satisfying the conditions $(U-V) \geq 1.1$ and $(B-V) \geq 0.8$; all the other objects were classed as {\it blue galaxies}. 

In fig. 1, top left, we show the resulting BH mass functions at $z=0$ from the aforementioned runs. Only objects located at the centers of galaxies were included. The solid curve  represents the BH mass function derived from data$^{23}$ and the grey zone represents the uncertainties. Simulated 
mass functions are similar but they overestimate slightly the BH density in comparison with the mass distribution calculated in Ref. 23. It is worth mentioning that the present simulations are only exploratory and they are not intended to reproduce exactly the observed properties of SMBH, but mainly to understand how the different feedback mechanisms and accretion modes affect the growth of seeds. 

\begin{figure}[h]
		\begin{tabular}{cc}
	\resizebox{59mm}{!}{\includegraphics[angle=270]{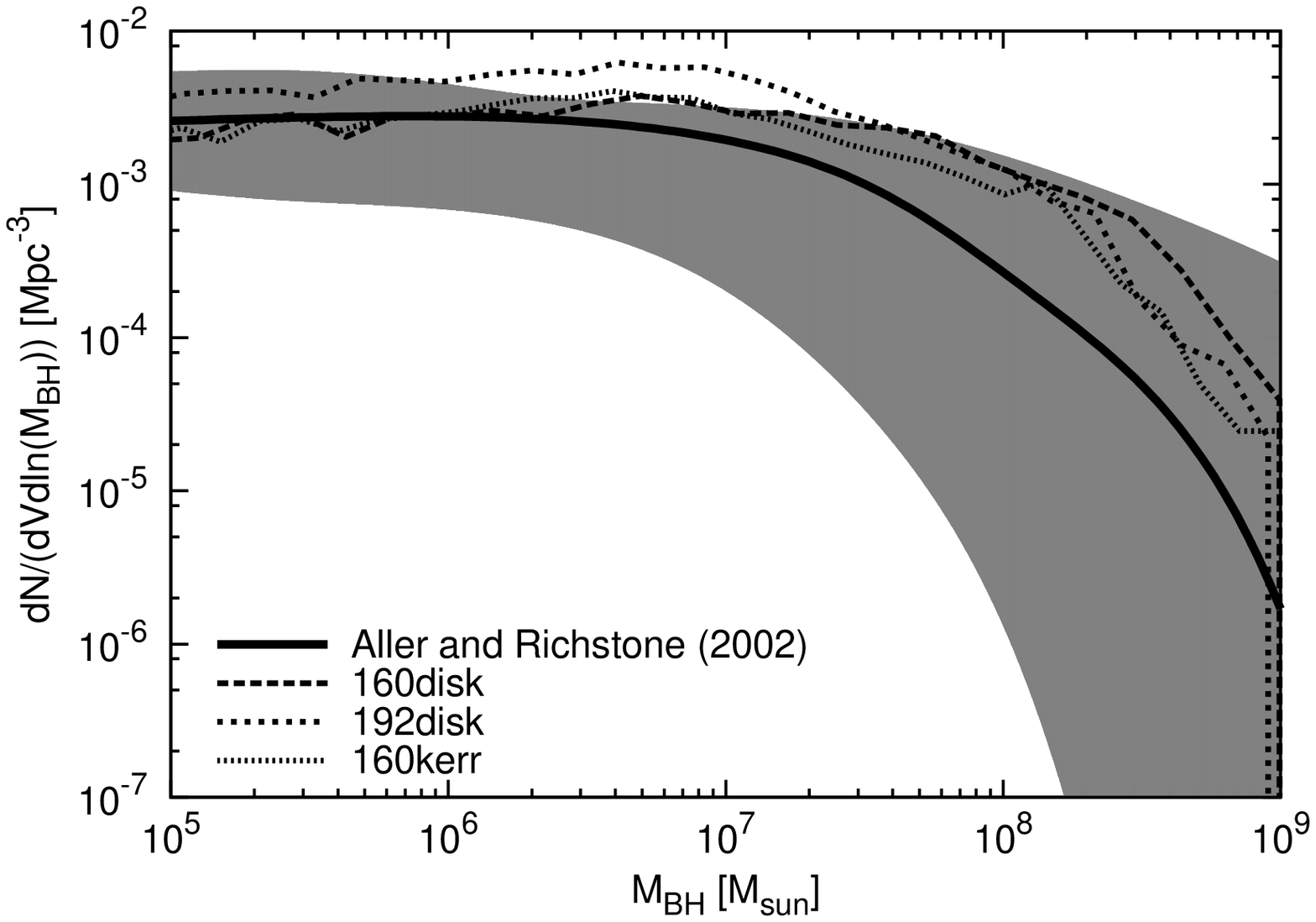}}&
	\resizebox{59mm}{!}{\includegraphics[angle=270]{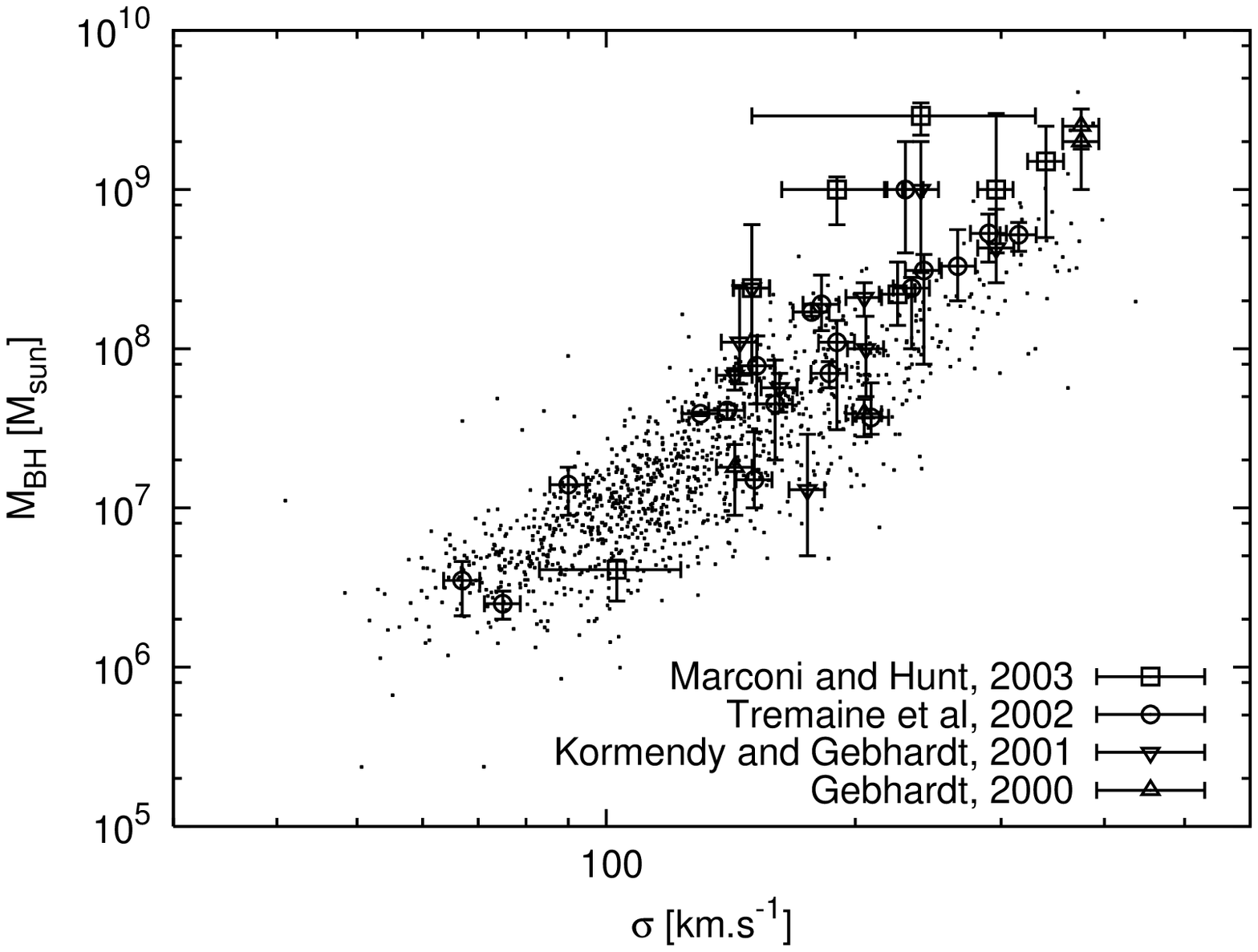}}\\
	\resizebox{59mm}{!}{\includegraphics[angle=270]{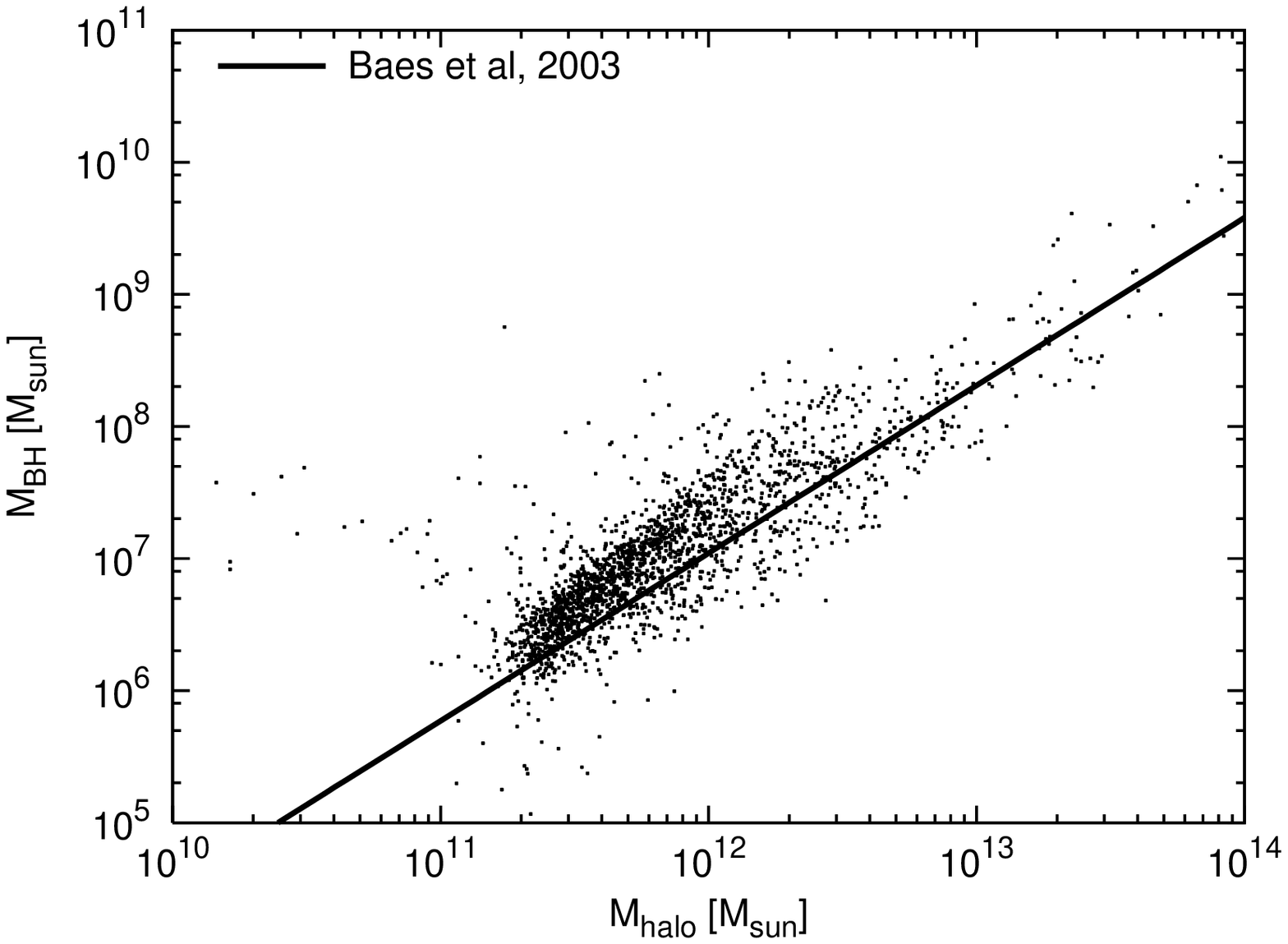}}&
	\resizebox{59mm}{!}{\includegraphics[angle=270]{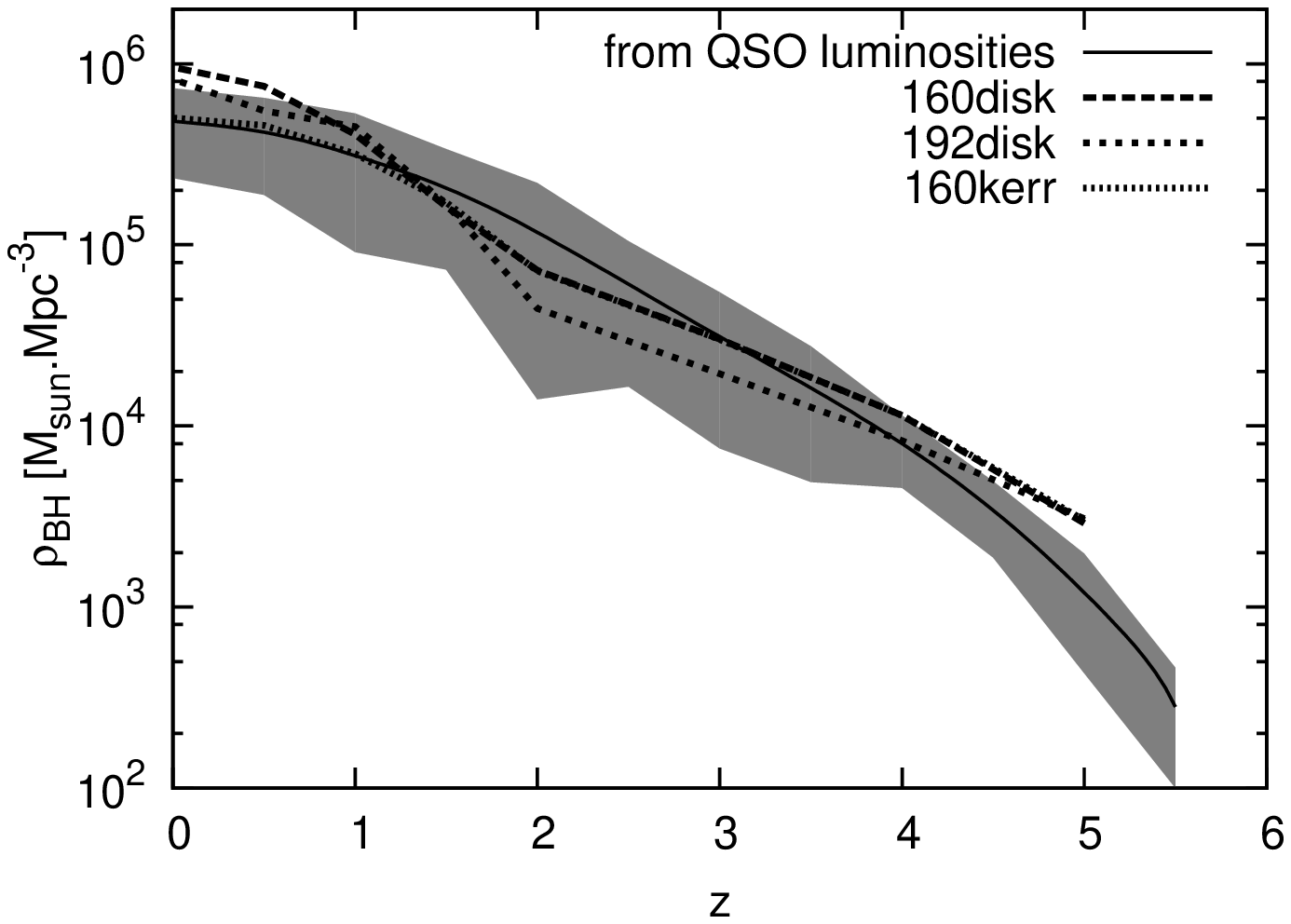}}\\
		\end{tabular}
	\caption{Top left: Simulated black hole mass functions in comparison with the distribution derived from data
(grey zone and solid line); top right: correlation between black hole masses and projected velocity dispersion (small
dots are simulated values and other symbols, observed data); bottom left: correlation between black hole masses and
dark halo masses (solid line, relation based on rotation velocities data); bottom right: evolution of the black hole
mass density in comparison with values derived from the bolometric luminosity function of QSOs.}
	\label{fig1}
\end{figure}

Not only the amount of energy injected either by supernovae and AGNs affect the resulting mass distribution but so also does the geometry of jets. The larger the opening angle, the larger is  the amount of gas blown out from the center of the  galaxy, quenching not only the star formation process in that region but also the growth of the central BH. The present results were obtained with an opening angle equal to $20^o$.

In fig. 1, top right, we show the $M_{BH}-\sigma$ relation at $z=0$. The agreement with the existing data is quite good, except for the BHs located in Cyg A, NGC 5252, NGC 3115 and NGC 4595, which seem to have masses higher than those expected from the observed velocity dispersion of their host galaxies. No significant differences are noticed between plots performed from the data of runs 160disk or 192disk. In fig. 1, bottom left, the relation between the BH mass and that of the host halo is shown. The solid line shown in this plot was derived in Ref. 10 from a combination of a fit representing the $M_{BH}-\sigma$ relation and another fit representing a relation between the velocity dispersion and the maximum
rotation velocity of the galaxy, an indicator of the dark halo mass. From our simulated data, we have obtained $M_{BH}\propto M_h^{1.06}$, a relation having a logarithmic  slope flatter than that derived in Ref. 16. Notice that the latter was derived at $z=1$ and not at $z=0$ as in the present work. 

Finally, in the bottom right of fig. 1, we show
the evolution of the BH mass density for the three  runs considered. For comparison, we also show the evolution of the BH mass density that we have derived from the bolometric luminosity function (BLF) of quasars given in Ref. 12. The agreement between the curves resulting from these three runs and that derived from the BLF of quasars is rather good, indicating that the average accretion rate history of BHs is well reproduced by our simulations. The present BH mass density estimated in Ref. 24 from existing data is $4.6^{1.9}_{-1.4}\times 10^5(h/0.7)^2~M_{\odot}Mpc^{-3}$.  From our simulations we derived values in the range $(5.0-9.6)\times 10^5(h/0.7)^2~M_{\odot}Mpc^{-3}$, consistent with observations. The lower value corresponds to  run 160kerr while the higher corresponds to run 160disk. The run 192disk gives an intermediate value corresponding to a BH mass density of $8.2\times 10^5(h/0.7)^2~M_{\odot}Mpc^{-3}$. It is interesting to see that a few percent  of the seeds, as a consequence of the hierarchical formation of structures, are not found in the centers  of galaxies but wandering in halos. These BHs have masses in the range $10^3-10^4~M_{\odot}$ since they were not able to grow. Most of them have presently small accretion rates  and are low-luminosity X-ray sources.

\section{Conclusions}

In this work, we report  our results on some properties of SMBHs resulting from cosmological simulations including the evolution of the host galaxies. BH seeds of $100~M_{\odot}$ inserted into proto-galaxies in the redshift range 15-20 are able to grow by gas accretion and coalescence, producing SMBH in the mass interval $10^{6-9}~M_{\odot}$, observed today in the centers of galaxies. Moreover,
our simulations also indicate that a population of intermediate mass ($10^{3-4}~M_{\odot}$) black holes is present in the halos of galaxies.

The SMBH mass function resulting from our simulations compares well with that derived from data, and the present SMBH mass density  derived from these exploratory numerical experiments is in the interval $(5-9)\times 10^5(h/0.7)^2~M_{\odot}Mpc^{-3}$, consistent with observations. It worth mentioning that better agreement could be achieved by fine tuning the feedback parameters, a procedure
that would be justified for higher resolution experiments. Another encouraging result is that at     
$z=0$, the mass of simulated BHs correlates well either with the velocity dispersion or the dark halo mass of the host galaxy.

In spite of these positive results, our simulations have also shown that around $z\sim 6$, SMBHs with masses  around $10^9~M_{\odot}$ have not yet formed. If the bright Sloan quasars observed at these redshifts are indeed so massive, their growth probably follows an ``anti-hierarchical" pattern, since they evolve faster then their host halos. Moreover, mildly supercritical Eddington accretion regimes are necessary to produce a quasi-exponential growth of seeds on a timescale of the order of 0.8 Gyr. These conditions could be found in rare massive halos that have already collapsed by that time.

A detailed study of the coalescence history of BHs could offer an opportunity to distinguish between different growth scenarios if future space experiments are able to detect the gravitational wave signal produced during the inspiral phase of these events.

\end{document}